\def\papertitle{Distortion Recovery: \\A Two-Stage Method for Guitar Effect Removal}
\def\paperauthorA{Ying-Shuo Lee$^*$}
\def\paperauthorB{Yueh-Po Peng$^*$}
\def\paperauthorC{Jui-Te Wu}
\def\paperauthorD{Ming Cheng}
\def\paperauthorE{Li Su}
\def\paperauthorF{Yi-Hsuan Yang}

\documentclass[twoside,a4paper]{article}
\usepackage{etoolbox}
\usepackage{dafx_24}
\usepackage{amsmath,amssymb,amsfonts,amsthm}
\usepackage{euscript}
\usepackage[T1]{fontenc}
\usepackage[utf8]{inputenc}
\usepackage{ifpdf}
\usepackage[english]{babel}
\usepackage{caption}
\usepackage{subfig} 
\usepackage{color}
\usepackage{booktabs}
\usepackage{tabularx} 

\input glyphtounicode
\ninept

\newcounter{numauth}
\newcounter{listcnt}
\newcommand\authcnt[1]{\ifdefined#1 \stepcounter{numauth} \fi}

\newcommand\addauth[1]{
\ifdefined#1 
\stepcounter{listcnt}
\ifnum \value{listcnt}<\value{numauth}
\appto\authorslist{, #1}
\else
\appto\authorslist{~and~#1}
\fi
\fi}
\authcnt{\paperauthorB}
\authcnt{\paperauthorC}
\authcnt{\paperauthorD}
\authcnt{\paperauthorE}
\authcnt{\paperauthorF}
\authcnt{\paperauthorG}
\authcnt{\paperauthorH}
\authcnt{\paperauthorI}
\authcnt{\paperauthorJ}
\def\authorslist{\paperauthorA}
\addauth{\paperauthorB}
\addauth{\paperauthorC}
\addauth{\paperauthorD}
\addauth{\paperauthorE}
\addauth{\paperauthorF}
\addauth{\paperauthorG}
\addauth{\paperauthorH}
\addauth{\paperauthorI}
\addauth{\paperauthorJ}

\usepackage{times}

\newif\ifpdf
\ifx\pdfoutput\relax
\else
   \ifcase\pdfoutput
      \pdffalse
   \else
      \pdftrue
\fi

\ifpdf 
  \usepackage[pdftex,
    pdftitle={\papertitle},
    pdfauthor={\authorslist},
    pdfsubject={Proceedings of the 27th International Conference on Digital Audio Effects (DAFx24)},
    colorlinks=false, 
    bookmarksnumbered, 
    pdfstartview=XYZ 
  ]{hyperref}
  \usepackage[pdftex]{graphicx}
\else 
  \usepackage[dvips]{epsfig,graphicx}
  \usepackage[dvips,
    pdftitle={\papertitle},
    pdfauthor={\authorslist},
    pdfsubject={Proceedings of the 27th International Conference on Digital Audio Effects (DAFx24)},
    colorlinks=false, 
    bookmarksnumbered, 
    pdfstartview=XYZ 
  ]{hyperref}
\fi
\usepackage[hypcap=true]{caption}
\title{\papertitle}

\sixaffiliations
{{\em \paperauthorA}\\{\href{https://web.ee.ntu.edu.tw/}{Department of Electrical Engineering} \\ National Taiwan University \\ Taipei, Taiwan \\ {\tt \href{mailto:r10921a16@ntu.edu.tw}{r10921a16@ntu.edu.tw}} }}
{{\em \paperauthorB}\\{\href{https://www.iis.sinica.edu.tw//}{Institute of Information Science} \\ Academia Sinica \\ Taipei, Taiwan \\ {\tt \href{mailto:yuehpo@iis.sinica.edu.tw}{yuehpo@iis.sinica.edu.tw}} }}
{{\em \paperauthorC}\\{\href{https://dafx2019.bcu.ac.uk/}{Positive Grid} \\ Henderson, USA \\ {\tt \href{mailto:ray.wu@positivegrid.com}{ray.wu@positivegrid.com}} }}
{{\em \paperauthorD}\\{\href{https://www.iis.sinica.edu.tw//}{Institute of Information Science} \\ Academia Sinica \\ Taipei, Taiwan \\ {\tt \href{mailto:hugowski@iis.sinica.edu.tw}{hugowski@iis.sinica.edu.tw}} }}
{{\em \paperauthorE}\\{\href{https://www.iis.sinica.edu.tw//}{Institute of Information Science} \\ Academia Sinica \\ Taipei, Taiwan \\ {\tt \href{mailto:lisu@iis.sinica.edu.tw}{lisu@iis.sinica.edu.tw}} }}
{{\em \paperauthorF}\\{\href{https://web.ee.ntu.edu.tw/}{Department of Electrical Engineering} \\ National Taiwan University \\ Taipei, Taiwan \\ {\tt \href{mailto:yhyangtw@ntu.edu.tw}{yhyangtw@ntu.edu.tw}} }}

\begin{document}
\ifpdf 
  \DeclareGraphicsExtensions{.png,.jpg,.pdf}
\else  
  \DeclareGraphicsExtensions{.eps}
\fi


\maketitle
\def\thefootnote{*}\footnotetext{These authors contributed equally to this work}\def\thefootnote{\arabic{footnote}}

\begin{abstract}
Removing audio effects from electric guitar recordings makes it easier for post-production and sound editing. An audio distortion recovery model not only improves the clarity of the guitar sounds but also opens up new opportunities for creative adjustments in mixing and mastering. While progress have been made in creating such models, previous efforts have largely focused on synthetic distortions that may be too simplistic to accurately capture the complexities seen in real-world recordings. 

In this paper, we tackle the task by using a dataset of guitar recordings rendered with commercial-grade audio effect VST plugins. 
Moreover, we introduce a novel two-stage methodology for audio distortion recovery. 
The idea is to firstly process the audio signal in the Mel-spectrogram domain in the first stage, and then use a neural vocoder to generate the pristine original guitar sound from the processed Mel-spectrogram in the second stage.
We report a set of experiments demonstrating the effectiveness of our approach over existing methods, through both subjective and objective evaluation metrics.
\end{abstract}

\section{Introduction}
\label{sec:intro}


Electric guitar effects such as distortion usually act as a decisive factor across various musical genres affecting the emotion, color, and the aesthetic taste of music.
For many music information retrieval (MIR) tasks, however, such guitar effects add another layer of complexity and can degrade the performance of MIR models.
For example, for automatic music transcription, Chen \textit{et al}. \cite{chen2022towards} found that guitar signals with different pedal effects negatively impact the accuracy of transcription. 
As such, \emph{distortion recovery}, the task of automatically removing effects from recorded tracks \emph{post hoc}, may provide a solution improving the performance of MIR models, including transcription,  source separation and automatic mixing systems \cite{martinez2021deep, steinmetz2021deep, martínezramírez2022automatic}.
For sound engineers, distortion recovery also make it easier for
sound editing.

In prior research, distortion recovery is usually treated as a special case of source separation or source enhancement \cite{imort_distortion_2022, rice2023general}. Specifically, it is assumed that the distorted signal can be represented as the sum of the clean signal and the ``effect signal'' (regarded as noise or another source). Signal processing and machine learning methods can then be developed to extract the clean signal from mixed or noisy ones \cite{defossez2021hybrid, choi2018phaseaware}. Although these methods have been shown to be promising,
we note that previous research mostly consider only synthetic distortions that may be too simplistic to reflect the complexities seen in real-world recordings.  
The complex and dynamic features of various effect pedals and Virtual Studio Technology (VST) plugins, combined with the diverse playing styles and recording environments, can all pose challenges for distortion recovery.
However, how such nuances in real-world recordings impact the performance of distortion recovery have not been studied thus far, to our best knowledge.



In this paper, we present two contributions to the task of distortion recovery.
Firstly, we propose a new technical approach that is inspired by recent advance in voice conversion and synthesis. Specifically, our approach contains two stages. In the first stage, we utilize a ``Mel denoiser'' which transforms the Mel-spectrogram of the distorted audio signal into that of the non-distorted, dry signal. In the second stage, we employ a 
neural vocoder to obtain the waveform of the dry signal. Experiments show that, compared to the prevalent single-stage approach, the proposed approach can better reinstate the intricate details inherent in the original guitar recording into the purified waveform. This preservation of the expressiveness and dynamic range of the original signal sets our method apart from the prior arts, demonstrating superior performance in auditory fidelity and processing efficiency. 


Secondly, 
we build and test the implemented models 
on two distinct datasets: one derived from software simulation using the Pedalboard \cite{sobot_peter_2023_7817838} as done in previous work, and the other from the ``BIAS FX2 ToneCloud presets''\footnote{\url{https://www.positivegrid.com/products/bias-fx-2}} using commercial-grade VST plugins released by a leading guitar amp and effect modeling company called Positive Grid.\footnote{\url{https://www.positivegrid.com/}}
This allows for a performance comparison of models in controlled versus real-world environments, offering new insights into the tested models.

Audio samples can be found at our demo page.\footnote{\url{https://y10ab1.github.io/guitar_effect_removal/}}
Moreover, as we use in-house data in our experiments (see Section \ref{sec:db}), for reproducibility we will create a dataset that can be publicly released (using the dry signals from the EGDB dataset \cite{chen2022towards}) and report evaluation result on that dataset on the demo page as well.

\section{Related Works}
\label{related_works}



The quality and clarity of audio signals usually play a crucial role in MIR applications such as music transcription and chord recognition. Chen \textit{et al}. \cite{chen2022towards} assessed the performance of guitar transcription using various settings, including dry Direct Input (DI) signal, and wet signals rendered with amps and real-world recordings sourced from YouTube, observing performance degradation on wet signals compared to the case of dry signals. 
Pauwels \textit{et al}. \cite{Pauwels201920YO} also showed 
that chord recognition datasets often feature clean and well-defined chords, which may not be practical in real-life situations, particularly in guitar signals where distortion effects are commonly used, calling for the need of  effect removal.



Distortion recovery is 
a novel and challenging task lying between the realms of signal processing and machine learning. 
Deep neural networks (DNNs) have been adopted for distortion recovery lately.
Imort \textit{et al}. \cite{imort_distortion_2022} explored the elimination of distortion and clipping from guitar tracks using various DNN architectures, finding that a model originally developed for source separation \cite{defossez2021hybrid} works the best. Their work signifies a pivotal shift towards employing deep learning for audio effect manipulation, suggesting the potential of DNNs in distinguishing and isolating the nuanced characteristics of distorted audio signals.

Expanding the scope to encompass general-purpose audio effect removal, Rice \textit{et al}. \cite{rice2023general} investigated a scenario with five specific audio effects: distortion, dynamic range compression, reverberation, chorus, and feedback delay. They devised a process named ``RemFX,'' which first detects whether a type of audio effect has been applied, and then removes each effect one at a time. They also showed that source separation-based model such as Demucs V3 \cite{defossez2021hybrid} and speech enhancement-based model such as DCUNet \cite{choi2018phaseaware} work well  for the removal of audio effects.
The goal of the present work is related to but different from theirs---we intend to build a model that removes combinations of multiple correlated distortion effects that are hard to be tackled individually.
However, as Demucs V3 and DCUNet have been shown promising in their setting, we adopt these two models as baselines in our experiments.

The neural vocoder, which employs neural networks to reconstruct waveforms from Mel-spectrograms, is widely used in audio processing. Modern neural vocoders, including MelGAN \cite{kumar2019melgan}, HiFiGAN \cite{kong2020hifi}, and iSTFTNet \cite{kaneko2022istftnet}, leverage generative adversarial networks (GANs) to achieve high-fidelity results that significantly surpass those of the traditional Griffin-Lim algorithm. In this study, we tackle the task of removing guitar distortion through a two-stage approach: initially processing the Mel-spectrogram, followed by employing a neural vocoder to convert it back into waveform. This method diverges from previous research, which predominantly performs distortion recovery in a single stage directly in the waveform domain (e.g., \cite{defossez2021hybrid, choi2018phaseaware}).


As for data, acquiring data from physical amplifiers presents significant challenges. Juvela \textit{et al}. \cite{juvela2023end} mechanized the process by attaching electric motors to each pertinent control of a physical amplifier, successfully gathering 4.5 hours of paired signals. However, the utilization of rudimentary algorithms \cite{imort_distortion_2022} and software applications such as Pedalboard \cite{rice2023general, sobot_peter_2023_7817838} might yield data that are not realistic enough.

\section{Method}
\label{method}
\subsection{Distortion Recovery Process}

The state-of-the-art techniques of audio distortion recovery \cite{imort_distortion_2022, rice2023general}, particularly concerning distortion effects, posit that the mixed signal, $y$, is represented as a linear blend of \textit{wet} signal $f(x)$ and the \textit{dry} signal $x$, where the nonlinear distortion function \( f(\cdot) \) is applied to \( x \):
\begin{equation}
y = \alpha f(x) + (1-\alpha)x\,,
\end{equation}
where $\alpha \in [0, 1]$ represents the influence of the distorted signal. This assumption basically stems from source separation and audio enhancement models \cite{ defossez2021hybrid, choi2018phaseaware, stoter2019open, chen2020dualpath}. 

Unlike the prior approach, we instead assume that  the distortion effect fundamentally alters the characteristics of the dry component such that it may not be identifiable within the processed output. Distortion typically generates highly nonlinear interactions, not merely attenuating but transforming the dry signal. More specifically, the mixed signal aforementioned can be approximated as a wet signal, $y$, expressed as follows:
\begin{equation}
\label{eq2}
y = f^*(x)\,.
\end{equation}
This realization prompts a shift from traditional linear models to a more sophisticated function that encapsulates the intricacies of this transformation. Inspired by voice conversion and synthesis \cite{van_Niekerk_2022,ren2019fastspeech,wu2022unified}, we articulate Equation \ref{eq2} as a two-stage model. The first stage focuses on recovering an approximation of the clean signal from the distorted wet signal, acknowledging that it might be devoid of certain fine details:
\begin{equation}
\hat{x}_\text{approx} = h(y)\,,
\end{equation}
where \( y \) is the distorted wet signal and \( h(\cdot) \) symbolizes the initial recovery function, striving to approximate the clean signal \( \hat{x}_\text{approx} \). Noting that \( \hat{x}_\text{approx} \) may lack the subtleties and nuances inherent to the original guitar signal, the second stage is designed to reinstate these details:
\begin{equation}
\hat{x} = g(\hat{x}_\text{approx})\,.
\end{equation}
Here, \( g(\cdot) \) is a refinement function tasked with restoring the finer characteristics and nuances to the estimated clean signal \( \hat{x}_\text{approx} \),  yielding the final restored signal \( \hat{x} \).

In the following subsections, we introduce the functions \( h(\cdot) \) and \( g(\cdot) \), referred to as the ``Mel Denoiser'' and ``Neural Vocoder,'' respectively. Together, they model a two-step restoration process. Initially, \( h(\cdot) \) approximates the clean signal from a heavily distorted (or wet) output. Following this, \( g(\cdot) \) refines the approximation, polishing it to achieve a high-fidelity restoration of the original dry guitar signal.

\subsection{Mel Denoiser: The First Stage of the Proposed Model}
\begin{figure}[t!]
\centering
\includegraphics[width=0.75\linewidth]{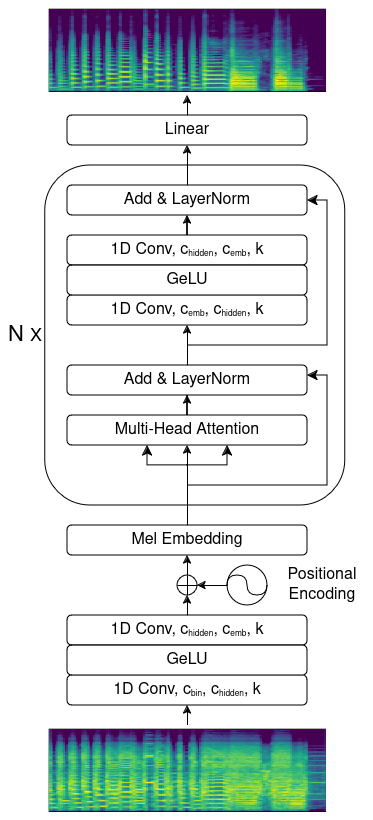} 
\caption{The architecture of the proposed Mel Denoiser. $N$ represents the number of layers, while $c_{bin}$, $c_{hidden}$, and $c_{emb}$ indicate the channel counts. The kernel size of the 1D convolution is denoted by $k$. Here, $c_{bin}$ matches the Mel- spectrogram bin count, and $c_{emb}$ corresponds to the embedding size. We configured $c_{hidden}$ to be four times larger than $c_{emb}$.} 
\label{fig:mel_denoiser_arch} 
\end{figure}

To initiate the denoising process, the wet waveform is first transformed into a Mel spectrogram. In our approach, each Mel spectrogram frame is treated as an embedding, effectively converting the Mel spectrogram into a sequence of embeddings. This conversion is ideal for Transformer-based architectures, which excel at processing sequences. However, traditional Transformers process the full-length sequence of hidden representations across all layers, resulting in high computational costs. Noting that adjacent frames are most significant for denoising tasks, and drawing inspiration from previous advancements in the text-to-speech (TTS) arena \cite{ren2019fastspeech}, we adopt a pure Transformer encoder with modifications as presented in Figure \ref{fig:mel_denoiser_arch}. Specifically, we replace the conventional feed-forward linear layer with two 1D convolution layers, incorporating a GELU activation function. This approach enhances the efficiency and effectiveness of the denoising process. Ultimately, the model is trained to generate a dry Mel spectrogram by removing the unwanted audio effects from the initial input.

\subsection{Neural Vocoder: The Second Stage of the Proposed Model}
Here we aim to convert the dry signal Mel spectrogram produced by the first stage into a dry waveform. In doing so, we employ the widely-used neural vocoder called HiFi-GAN \cite{su2020hifi}, which leverages generative adversarial networks for waveform generation. The generator within HiFi-GAN takes the Mel spectrogram as input and employs transposed convolution layers to progressively upsample the signal until the length of the output matches that of the target waveform. HiFi-GAN features two discriminators: the multi-period discriminator and the multi-scale discriminator. The former is designed to capture various periodic components of the raw waveforms, while the latter focuses on identifying patterns across different lengths of the raw waveforms, ensuring a rich and accurate audio reproduction.

\section{Experiments}
\label{experiments}

\subsection{Experimental setup}
\label{sec:exp_setup}
The audio signals are sampled at 44.1 kHz. The Mel-spectrogram is configured with 128 bins, with an window size 
of 2,048, and a hop length of 512. The Mel Denoiser comprises 12 blocks, each including a self-attention layer with an embedding size of 384, and employs 1D convolutional kernels with size of [9, 1]. For optimization, we use AdamW with a learning rate of 1e--5, a learning rate decay of 0.999999 at each step, and a batch size of 64. The HiFi-GAN implementation is sourced from the vtuber-plan project.\footnote{\url{https://github.com/vtuber-plan/hifi-gan}} The HiFi-GAN is trained with a modified LS-GAN loss, L1 Mel-spectrogram loss, and feature matching loss identical to the original paper, while the Mel Denoiser is only trained with L1 Mel-Spectrogram Loss. Our training regimen begins with separate training of the Mel Denoiser and the Neural Vocoder, conducting 1.5 million steps for the Mel Denoiser and 1 million steps for HiFi-GAN, followed by a ``fine-tuning'' phase for HiFi-GAN using the output from the Mel Denoiser for an additional 0.5 million steps. All experiments were conducted on a single RTX 4090 GPU.

\subsection{Baseline Models}

To benchmark the effectiveness of our proposed model in the domain of audio distortion recovery, we draw comparisons with three notable models renowned for their contributions to audio processing tasks: Demucs V3 \cite{defossez2021hybrid}, DCUnet \cite{choi2018phaseaware}, and HiFi-GAN Denoiser \cite{su2020hifi}. These models are selected for their relevance and demonstrated success in tasks closely aligned with our objectives, such as source separation, denoising, and audio enhancement.

\textbf{Demucs V3}, a.k.a., Hybrid Demucs, is an extension of the U-Net model designed for musical source separation \cite{defossez2021hybrid}. It combines convolutional layers with LSTM units to capture audio signal dependencies across different scales. 
It has been adopted in RemFX \cite{rice2023general} for distortion removal.

\textbf{DCUnet} is another variant of the U-Net architecture that is designed to work with complex spectrograms by employing complex convolutions. It excels in tasks requiring detailed spectral manipulation, such as speech enhancement and audio denoising, due to its capacity to preserve intricate phase and magnitude information. It has also been adopted in RemFX \cite{rice2023general}.

\textbf{HiFi-GAN Denoiser:} Distinct from the previously mentioned HiFi-GAN neural vocoder \cite{kong2020hifi}, the HiFi-GAN denoiser operates on a waveform-to-waveform basis, targeting the elimination of a wide array of noises, reverberations, and equalization distortions present in recordings. This denoiser utilizes a feed-forward WaveNet architecture, complemented by discriminators that operate on both the waveform and Mel- spectrogram scales. Additionally, it extracts deep features from the discriminators to enhance the perceptual quality of the audio output. This approach ensures the denoised audio maintains a high level of clarity and fidelity, making the HiFi-GAN Denoiser a strong competitor for our model, especially in terms of maintaining audio quality while removing distortion effects.

\textbf{Ours-Base:} In addition to the set up mentioned in Section \ref{sec:exp_setup}, we trained a variant of our model with fewer trainable parameters for comparison. This base model adopts all the configurations, except it scales down the number of layers from 12 to 8 and reduces the embedding size from 384 to 256. 

All the implemented models were trained from scratch for 1.5 million steps using one of the datasets described in Section \ref{sec:db}, with early stopping to halt training if no advancement in performance was observed. The AdamW optimizer, with a learning rate of 1e--4, facilitated the optimization process. Batch sizes were selected to optimize memory usage and computational efficiency, given the diverse memory requirements across models. The primary training objective for all models was the minimization of the L1 loss on the waveform. Notably, for the HiFi-GAN Denoiser \cite{su2020hifi}, the training also involved L2 loss on log spectrograms and additional adversarial and deep feature matching losses. These losses are particularly effective at capturing and improving perceptual aspects of audio quality, aiming for a denoised output that  resembles the natural properties of clean speech signals.

\subsection{Datasets}
\label{sec:db}

We consider two datasets in our experiments. 


\textbf{VST-derived Data:} 
To have a broad and varied dataset of paired signals, we utilize an in-house dataset containing 80 hours of electric guitar dry and wet signal pairs, provided by Positive Grid. 
Each clip lasts 4 seconds and is sampled at 44.1 kHz.
This dataset complies with Positive Grid's privacy policy and GDPR guidelines, ensuring the protection of personal data and user privacy. 
Specifically, the dry signals are contributed by 14 professional guitarists under consent agreements, while the wet signals are produced using the BIAS FX2 VST plugins of Positive Grid, with presets randomly selected from its ``ToneCloud'' library, which includes over 90,000 options. 
In general, VST plugins often consist of multiple stages of signal processing, including preamp modeling, tone shaping, cabinet simulation, effects processing, and post-pro\-cessing. Each stage adds complexity to the signal chain and contributes to the overall sound. 
To focus on distortion-related effects, we consider noise gate, EQ, compressor, overdrive, distortion, and amplifier, while other effects such as delay, modulation, reverb, and pitch shifting are excluded. 
Notably, while our dataset more accurately reflects typical guitarist recording conditions, discrepancies remain when compared to various real-world sources, such as YouTube recordings captured in different environments.

\textbf{Synthetic Distortion Data:} To enable comparison with prior research, we also created synthetic distortion effects using the Pedalboard \cite{sobot_peter_2023_7817838} on the same dry signals of the previous VST-derived data. Distortion and clipping effects were randomly applied to the dry signals. The gain levels $\gamma$ for the distortion effect were uniformly chosen from $\gamma \in [20, 50]$ dB, and the clipping thresholds $\tau$ were uniformly selected from $\tau \in [-50, -20]$ dB.

\subsection{Objective Evaluation Metrics}



We consider the following metrics to quantitatively measure the performance of the implemented models.

\textbf{Fréchet Audio Distance (FAD)} \cite{kilgour2019frechet}: Inspired by the Fréchet Inception Distance (FID) used in computer vision, FAD  measures the similarity between distributions of real and synthesized audio features as extracted by a deep learning model. A lower FAD score suggests a higher resemblance to the target audio distribution, indicative of superior audio quality. We report FAD score calculated by pretrained CLAP model \cite{laionclap2023}.
    
\textbf{Error-to-Signal Ratio (ESR)} \cite{esrloss} is a metric that quantifies the proportion of the error signal relative to the desired signal. It provides an insightful measure of the distortion or unwanted components present in the output audio. Lower ESR values indicate higher fidelity in signal reconstruction. 

\begin{table*}[ht]
\centering
\begin{tabular}{@{}lr|cccc|cc@{}}  
\toprule
                 & Params.           & FAD $\downarrow$ & ESR $\downarrow$ & SISDR $\uparrow$ & MR-STFT $\downarrow$ & AQ $\uparrow$ & DL $\uparrow$ \\ \midrule  
Demucs V3~\cite{defossez2021hybrid}        & 83.5M             & 0.383            & \textbf{0.869}   & ~~2.984            & 2.236                & 1.66$\scriptstyle{\pm0.83}$     & 1.86$\scriptstyle{\pm0.96}$     \\
DCUnet~\cite{choi2018phaseaware}           & 7.7M              & 0.249            & 0.968            & ~~4.085            & 1.821                & 2.10$\scriptstyle{\pm1.11}$     & 1.99$\scriptstyle{\pm1.05}$     \\
HifiGAN denoiser~\cite{su2020hifi} & \textbf{1.3M}     & 0.224            & 1.216            & ~~6.212            & 2.271                & 2.67$\scriptstyle{\pm1.12}$     & 2.77$\scriptstyle{\pm1.26}$     \\ 
\midrule
Ours-Base        & 45.9M          & 0.083             & 2.808            & 27.608           & 1.568              &  ---         & ---         \\ 

Ours-Large       & 101.7M            & \textbf{0.080}   & 2.290            & \textbf{28.650}  & \textbf{1.419}       & \textbf{3.54}$\scriptstyle{\pm1.10}$     & \textbf{3.86}$\scriptstyle{\pm1.08}$     \\ 
\midrule
Ground Truth     &   ---               &   ---              &   ---              &   ---              &   ---                  & 4.25$\scriptstyle{\pm0.89}$     & 4.29$\scriptstyle{\pm0.95}$     \\ 
\bottomrule
\end{tabular}
\caption{The number of parameters and performance of various models trained on VST-derived data.  Ours-Large leads to the best result across various objective (middle) and subjective (rightmost) metrics, while having a similar number of parameters as Demucs V3. Ground Truth serves as the high anchor for the two subjective evaluation metrics 
AQ (audio quality) and DL (dryness level). The arrows $\uparrow$ and $\downarrow$ indicate the higher or the lower the better; best result in each column is highlighted in bold.}
\label{tab:audio_quality_comparison}
\end{table*}
    
\textbf{Scale-Invariant Signal-to-Distortion Ratio (SI-SDR)} serves as a robust alternative to the traditional Signal-to-Distortion Ratio (SDR) \cite{LeRoux2018SDRH} by providing a scale-invariant measure of signal quality. It is particularly adept at evaluating the degree of distortion removal and the fidelity of the signal reconstruction. Higher SI-SDR values correlate with less audible distortion, suggesting a superior perceptual quality of the recovered audio signal. 

\textbf{Multiresolution STFT (MR-STFT)} \cite{yamamoto2020parallel, SteinmetzauralossAL}: In contrast to the conventional STFT where a singular trade-off between time and frequency resolution is inevitable, MR-STFT employs multiple STFT analyses with varied window sizes and resolutions. Doing so enables MR-STFT to capture both fine-grained temporal details and broad frequency characteristics within the same audio signal. Therefore, MR-STFT offers insight into the intricate time-frequency attributes of audio signals. 

\textbf{Number of Parameters:} This metric reflects the total number of trainable parameters within the model. A model with fewer parameters is generally more efficient, with a reduced memory footprint and faster inference capabilities.

\subsection{Subjective Evaluation Metrics}

Subjective evaluations provide a critical measure of an audio processing model's performance, offering insights that objective metrics cannot capture. In this assessment, Mean Opinion Scores (MOS) are employed, where 26 professional guitarists and music producers critically evaluate the perceptual quality of audio and the proficiency of models in restoring distortion-affected signals. To facilitate this, 10 distinct sets of audio samples, not included in the training set, were curated for auditory evaluation. Each set includes five audio files: the unprocessed original \textit{Dry} signal and the outputs processed by Demucs V3, DCUNet, HifiGAN-denoiser, and our model. The participants appraise each audio file, focusing on the quality and the effectiveness of distortion mitigation, on a scale from 1 to 5 (the higher the better).

\textbf{Audio Quality (AQ):} The AQ metric encompasses the overall sound quality after distortion recovery. It reflects the listeners' perception of clarity, fidelity, and the absence of unwanted artifacts. A higher MOS in AQ signifies that listeners perceived the audio as high quality, suggesting effective recovery of the original signal.

\textbf{Dryness Level (DL):} The DL metric evaluates the extent to which the effects, particularly distortion, have been removed from the guitar signal. A higher MOS in DL indicates a signal that listeners perceive as closely resembling the original, unaffected dry sound, implying that the corresponding model successfully removes the intended effects, restoring the natural state of the signal as heard by the listeners.


\section{Results}
\label{results}
\subsection{Audio Quality and Model Efficiency}
To conduct a comprehensive comparison of all models in a realistic scenario, we trained all models with the VST-derived data. Table \ref{tab:audio_quality_comparison} presents a comparative evaluation of audio quality metrics.
We see that the proposed model (i.e., Ours-Large) achieves the lowest FAD score, indicating alignment with the true distribution of the VST-derived data. Moreover, it secures the highest SI-SDR value, reflecting exceptional ability in signal reconstruction fidelity. However, possibly because our model is less sensitive to phase variations, it does not score well on ESR. We note that the ESR metric may not always align with subjective metrics that reflect human perception of music quality. 
Further research could be beneficial to evaluate the alignment of ESR metrics with human perceptual quality.
Demucs V3, despite achieving the lowest ESR score, showing fewer errors in the output signal, falls short compared to our model in other aspects. DCUnet, while computationally efficient, lags behind in performance. The fewer parameters of HiFi-GAN Denoiser hint at a trade-off in audio quality when compared to our more sophisticated model. Our-Base exhibit strong performance with few parameters, with Ours-Large showing exceptional proficiency across all evaluated metrics.

\begin{figure*}[htbp]
\centering
\includegraphics[width=\linewidth]{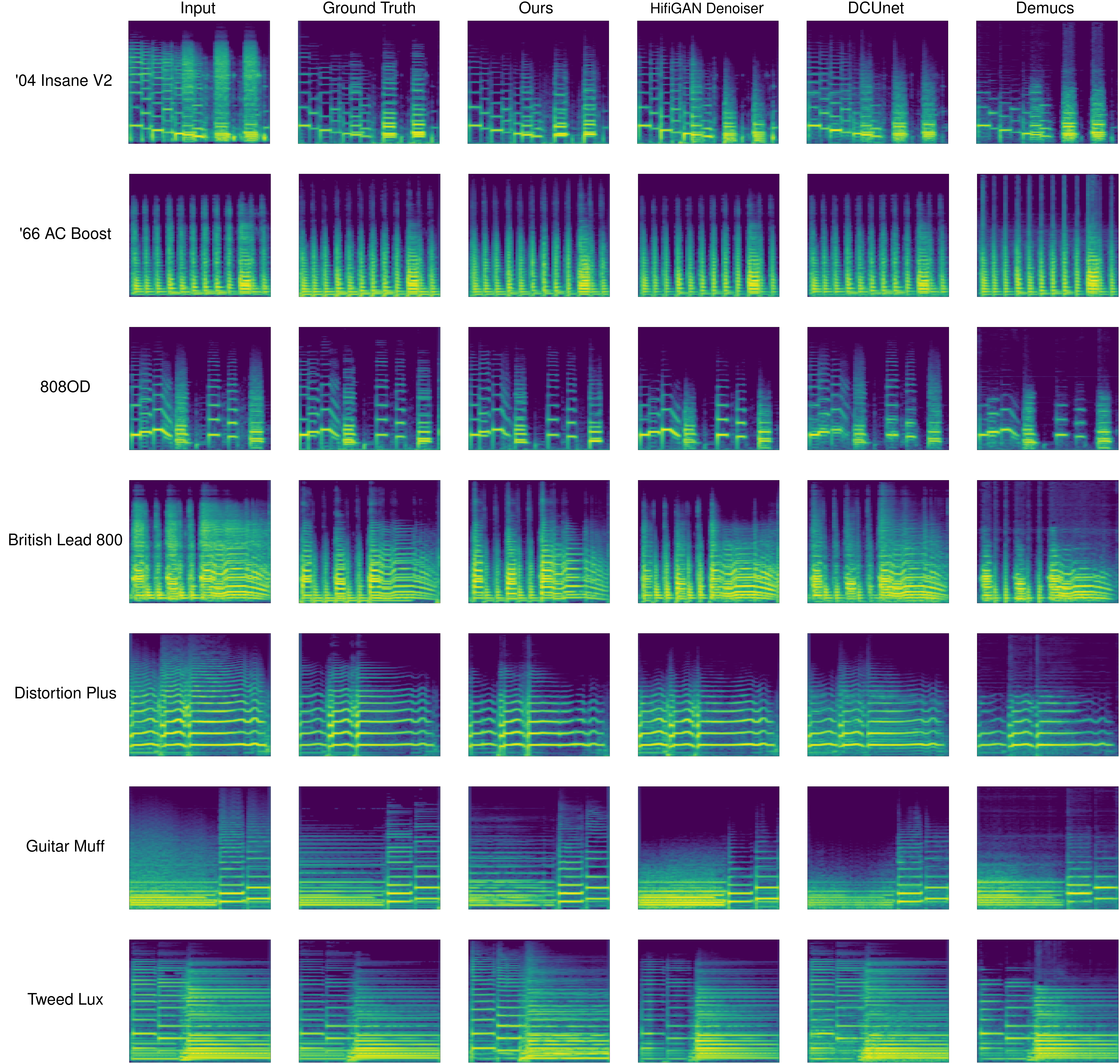} 
\caption{The Mel-spectrograms of the input wet signal, target dry signal, along with the output of the proposed model, the HiFiGAN Denoiser \cite{su2020hifi}, DCUnet \cite{choi2018phaseaware}, and Demucs V3 \cite{defossez2021hybrid}, across a total of seven different VST plugin effects. Our model demonstrates a closer resemblance to the target signal, showcasing superior distortion reduction capabilities and better preservation of overtone characteristics.} 
\label{fig:mel_compare} 
\end{figure*}

Additionally, we present the Mel-spectrogram of several examples in Figure \ref{fig:mel_compare}. Our model closely matches the target while baseline models fail to retain high-frequency signals and do not effectively eliminate noise.

\begin{table}[htbp]
\centering
\resizebox{\linewidth}{!}{
\begin{tabular}{@{}lcccc@{}}
\toprule
                          & FAD $\downarrow$   & ESR $\downarrow$   & SI-SDR $\uparrow$ & MR-STFT $\downarrow$ \\ 
\midrule
Demucs V3 (Synthetic) & \textbf{0.375}  & 2.436          & 16.792          & 3.589\\
DCUnet (Synthetic)    & 0.392           & \textbf{1.002} & 16.539          & 2.856\\ 
Ours (Synthetic)      & 0.455           & 1.827          & \textbf{31.790} & \textbf{2.170}\\
\midrule
Demucs V3 (VST)      & 0.383           & \textbf{0.869} & ~~2.984           & 2.236 \\
DCUnet (VST)             & 0.249           & 0.968          & ~~4.085           & 1.821 \\ 
Ours (VST)  & \textbf{0.080} & 2.290          & \textbf{28.650} & \textbf{1.419}\\
\bottomrule
\end{tabular}
}
\caption{Performance comparison of models trained with VST-derived data (VST) and synthetic distortion data (Synthetic). The table highlights our model's superior performance in terms of FAD, SI-SDR, and MR-STFT metrics when trained exclusively on the VST-derived data, in contrast to Demucs V3 and DCUnet, which exhibit lower performance across both types of datasets. This underlines our model's robustness and its enhanced capability to approximate the target dry signal accurately in complicated real-world VST-derived acoustic settings.}
\label{tab:real_vs_synthetic}
\end{table}

\subsection{Training with the VST-derived Data~vs.~with the Synthetic Distortion Data}

The analysis of distortion recovery unfolds in two  phases: initially with the synthetic data and later with the VST-derived data.
These datasets are instrumental in assessing the versatility of model and their capacity to adapt to varying acoustic environments.

Upon training the models with the synthetic data and evaluating them on the VST-derived data, we find that none of the models reach desirable outcomes in terms of the FAD. However, they do achieve relatively high SI-SDR scores. This juxtaposition suggests that with the synthetic data, while fostering high SI-SDR performance, falls short in preparing models for the nuanced complexities encountered in the VST-derived data, as reflected by the elevated FAD values. The FAD metric reveals a significant gap between the model outputs and the practical data, highlighting a potential limitation of the synthetic training data in replicating the full spectrum of the VST-derived audio nuances.

Furthermore, when the models are trained with the VST-derived data, our model exhibits a marked improvement over the other models across most metrics. Notably, our model achieves a dramatically lower FAD score (0.080 compared to 0.249 and 0.383 by DCUnet and Demucs V3, respectively) and a significantly higher SI-SDR value (28.650 compared to 4.085 and 2.984 by DCUnet and Demucs V3, respectively) when evaluated on VST data. These results suggest that our model is particularly adept at handling the diverse and complex nature of the VST-derived audio signals, providing a more accurate and reliable removal of effects.

The superior performance of our model on the VST-derived data underlines the importance of training with data that closely mimics the target environment. It also supports the idea that models trained on more representative audio signals are more likely to generalize well to real-world scenarios,\footnote{Additional experiments conducted on the EGDB dataset \cite{chen2022towards} are available on the demo page. The dataset and results can be accessed at \url{https://y10ab1.github.io/guitar_effect_removal/}.} confirming the efficacy of our model in managing the unpredictable variations present in realistic audio environments.

\begin{figure}[t]
\centering
\includegraphics[width=\linewidth]
{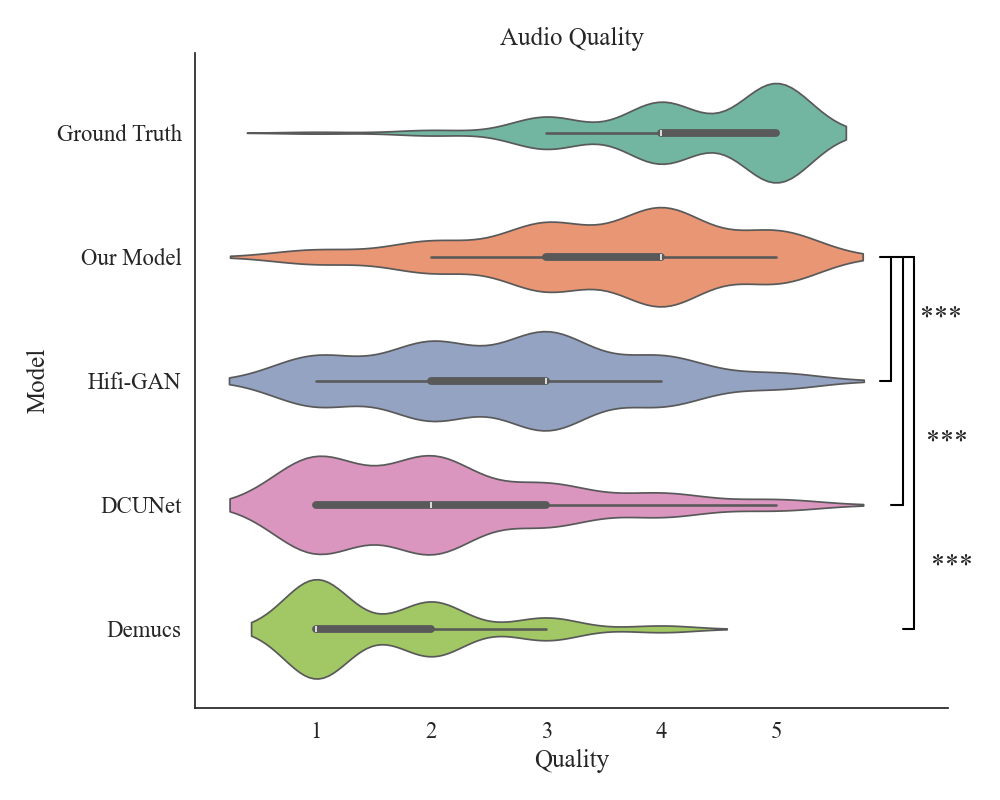} 
\caption{Mean Opinion Scores for Audio Quality (AQ). The distribution indicates that our model primarily achieved ratings around 4 points, signifying a high level of signal quality post-distortion recovery. ($\text{***}=p<.001$ in statistical test).} 
\label{fig:aq} 
\end{figure}
\begin{figure}[t]
\centering
\includegraphics[width=\linewidth]{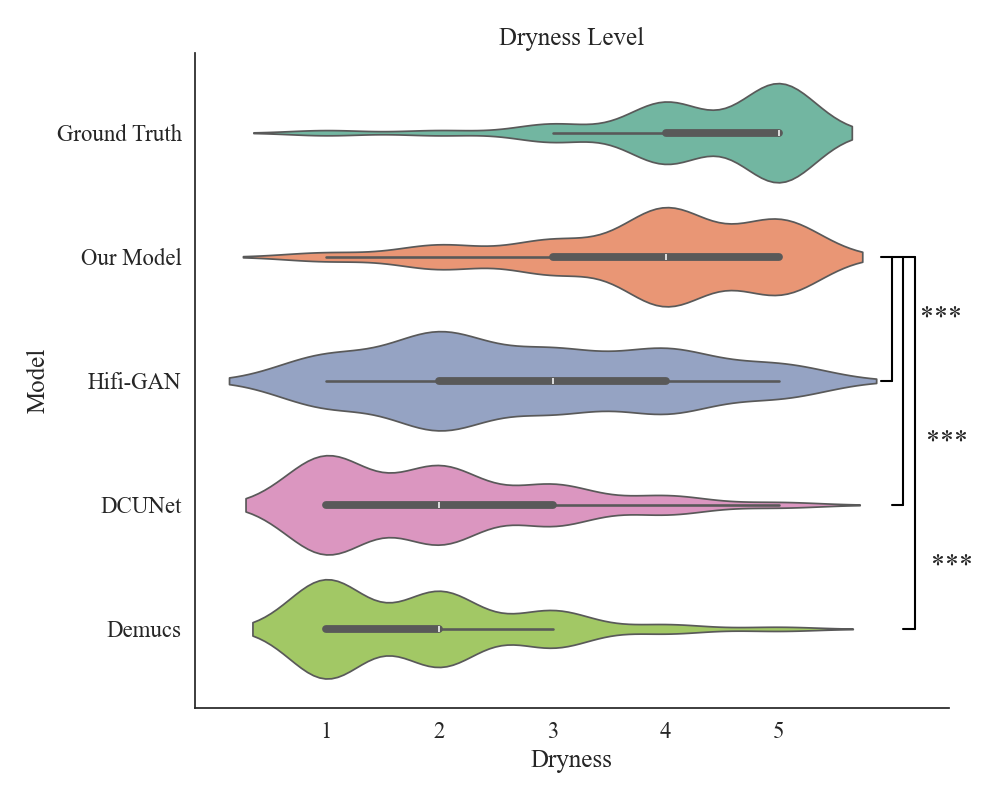} 
\caption{Mean Opinion Scores for Dryness Level (DL). The concentration of ratings around 4 points for our model suggests the dryness of recovered signal is favorably compared to the ground truth, demonstrating effective distortion removal. ($\text{***} = p < .001$.)} 
\label{fig:dl} 
\end{figure}

\subsection{Subjective Quality Evaluations}
Audio quality and dryness level play a crucial role in assessing the performance of distortion recovery models, offering insights into the perceived quality from the listener's perspective. To gauge the effectiveness of our model compared to existing baselines, we conducted a user study focusing on these two aspects.

Figure \ref{fig:aq} shows a violin plot of audio quality (AQ) ratings for various models. Our model predominantly received ratings around 4 points, indicating listeners highly regard the recovered signal’s quality. This consistent high rating sets a benchmark in distortion recovery. Similarly, Figure \ref{fig:dl} displays ratings for the dryness level (DL) of audio signals, reflecting how well the recovered signal matches a clean, undistorted ground truth. The violin plot indicates our model frequently scores 4, demonstrating its effectiveness in removing distortion and maintaining the signal’s natural features.
Post-hoc analyses using the Tukey HSD test following the ANOVA demonstrated  significant differences in MOS ratings between the models. The statistical findings are consistent with objective metrics, showing that our model outperforms Hifi-GAN, DCUNet, and Demucs in terms of audio quality and dryness (all with $p\text{-value} < 0.001$). 

\begin{table}[]
\centering
\begin{tabular}{@{}lcccc@{}}
\toprule
                  & FAD $\downarrow$  & ESR $\downarrow$  & SI-SDR $\uparrow$  & MR-STFT $\downarrow$\\ 
\midrule
Ours-Base              & 0.128          & \textbf{1.842} & 27.532          & 1.430 \\
 + finetune   & 0.083          & 2.808          & 27.608          & 1.568  \\ 
\midrule
Ours-Large             & 0.129          & 1.976          & 27.802          & \textbf{1.358} \\
 + finetune  & \textbf{0.080} & 2.293          & \textbf{28.651} & 1.419\\ 
\bottomrule
\end{tabular}
\caption{Ablation study comparing different model sizes and the effect of vocoder fine-tuning. See Section \ref{sec:ablation} for discussions.}
\label{tab:ablation_study}
\end{table}

\subsection{Model Architecture Ablation}
\label{sec:ablation}

Finally, we conducted an ablation study to examine the effects of varying model sizes and the impact of the fine-tuning of the vocoder as described in Section \ref{sec:exp_setup}. According to our informal subjective listening, the larger model (Ours-Large) produces outputs more closely resembling the target, particularly in polyphonic compositions compared to the base model. Additionally, fine-tuning of the vocoder helps in reducing artifacts, yielding outputs that are more realistic from a human perspective. However, the larger model does not outperform the base model in objective metrics; in fact, fine-tuning of the vocoder appears to worsen the ESR and MR-STFT scores. We argue that these metrics may not fully capture human perceptions of sound quality.

\section{Conclusion}
\label{conclusion}
In this paper, we have presented a two-stage methodology for removing audio effects from electric guitar tracks, significantly advancing the state-of-the-art for effect recovery.
Leveraging a novel approach that combines Mel-spectrogram purification with neural vocoder-based reconstruction, our model outperforms existing ones in producing high-fidelity original sounds from distorted guitar recordings.
Moreover, through a comprehensive evaluation employing a broad mix of VST plugins, we have shown that the proposed model 
performs well not only for simplistic distortion effects tested in prior works, but also for more complicated VST-derived effects that have not been well studied before. 

In future work, we plan to extend our approach to more challenging real-world settings, e.g., on guitar recordings sourced from YouTube.
It would also be interesting to apply our model to downstream tasks such as guitar transcription and effect modeling.


\bibliographystyle{IEEEbib}
\bibliography{DAFx24_tmpl} 


\end{document}